# Modal Decompositions of the Kinematics of Crevalle Jack and the Fluid-Caudal Fin Interaction

Muhammad Saif Ullah Khalid[1,2], Junshi Wang[3], Imran Akhtar[4], Haibo Dong[3], Moubin Liu[1,2*]

[1] Institute of Ocean Research, Peking University, Beijing, People's Republic of China
[2] Key State Laboratory of Turbulence and Complex Systems, Department of Mechanics and Engineering Science, Peking University, Beijing, People's Republic of China
[3] Department of Mechanical and Aerospace Engineering, University of Virginia, Charlottesville, 22904, VA, USA
[4] Department of Mechanical Engineering, NUST College of Electrical & Mechanical Engineering, National University of Sciences & Technology, Islamabad, 44000, Pakistan

* Corresponding Author, Email: mbliu@pku.edu.cn

To understand the governing mechanisms of bio-inspired swimming has always been challenging due to intense interactions between the flexible bodies of natural aquatic species and water around them. In this paper, we employ advanced modal decomposition techniques; proper orthogonal decomposition and dynamic mode decomposition, to extract energetically strongest spatio-temporal orthonormal components of complex kinematics of a Crevalle Jack (*Caranx hippos*) fish. Then, we present a computational framework for handling fluid-structure interaction related problems in order to investigate their contributions towards the overall dynamics of highly nonlinear systems. We find that the undulating motion of this fish can be described by only two standing-wave like spatially orthonormal modes. Constructing the data set from our numerical simulations for flows over the membranous caudal fin of the Jack fish, our modal analyses reveal that only the first few modes receive energy from both the fluid and structure, but the contribution of fluid in the higher modes is minimal. For the viscous and transitional flow conditions considered here, both spatially and temporally orthonormal modes show strikingly similar coherent flow structures. Our investigations are expected to assist in developing data-driven reduced-dimensional mathematical models to examine the dynamics of bio-inspired swimming robots and develop new and effective control strategies to bring their performance closer to real fish species.

## 1. INTRODUCTION

For the last two decades, the scientific community has made a lot of progress to understand the natural aquatic locomotion by numerous species that could enable them to utilize the discovered hydrodynamic mechanisms to propose efficient and maneuverable designs for bio-inspired underwater vehicles (Fish, 2020). Despite this substantial amount of efforts, there exists a huge gap to design effective flow control strategies. A major difficulty in this pursuit relates to the involvement of uncertain real conditions in large water reservoirs, such as oceans and rivers, to be faced by swimming robots and the prediction of their dynamical states impacted by numerous physical factors. In this context, data-driven techniques come up as great candidates for predicting complex nonlinear flow dynamics and tuning the kinematics of the flexible body-structures of fish-like robots [ (Brunton, et al., 2020; Verma, et al., 2018).  This scenario has also raised the requirement to develop effective reduced- or low-order mathematical models to describe the mechanics of these engineering systems that would open up new horizons to apply machine learning or deep learning control techniques in this field. For the dimensionality reduction, advanced modal decomposition techniques, such as proper orthogonal decomposition (POD), dynamic mode decomposition (DMD) and their variants (Rowley & Dawson, 2017) provide great tools to extract primary features of nonlinear dynamical systems without really solving the governing equations. Previously, several people reported their efforts to utilize these techniques to understand underlying hydrodynamic mechanisms for bio-inspired flows. Ting and Yang (2009) used singular-value decomposition (SVD) method to extract key flow features in the two-dimensional wake of a fish. Some other studies (Liang & Dong, 2015; Li, et al., 2016; Li, et al., 2017; Han, et al., 2017) presented the utility of POD technique based on a traditional eigenvalue decomposition analysis for flows around flapping wings and plates. These investigations also proposed the concept of virtual force to classify the modes to find their contributions in the production of life and thrust forces.

There is an interesting way of investigating the effects of dominant structural modes to quantify their relative contributions in the production of total hydrodynamic forces on bio-inspired structures during their steady swimming. For example, Bozkurttas et al. (2009) utilized the SVD formulation to determine that only three structural POD modes were sufficient enough to model the complex dynamics of a flexible pectoral fin of a bluegill sunfish. They concluded that the kinematics reconstructed by the mean and three oscillatory POD modes was able to produce 92% of the thrust force generated by the full-order kinematics of the pectoral fin. Ren and Dong (2016) utilized a similar methodology to decompose the morphing wing kinematics of a hovering dragonfly to examine the effects of the POD modes on its aerodynamic performance.

Besides, there were a few recent efforts to break down the travelling-wave like motion of different carangiform swimmers (Feeny & Feeny, 2013; Tanha, 2018). Feeny and Feeny (2013) considered the transverse kinematics of a Whiting and carried out complex modal analysis. They found that a single complex mode was enough to represent the transverse wavy motion of the fish. In this formulation, this complex mode had two components out of which the real one showed a standing wave and the imaginary part represented the traveling wave like structure. Following a similar approach, Tanha (2018) employed the modal information to approximate important kinematic parameters, such as oscillation amplitudes and phases and their dependence on time and spatial location of fish bodies. However, the low-dimensional analyses conducted by the afore-mentioned studies were limited to the kinematics of flexible structures. The connections between dominant kinematic modes and primary flow features of complex fluid-structure interaction-based systems are still elusive.

Recent advancements in the field of modal analysis to characterize complex dynamical systems have opened doors to analyze the underlying mechanics of bio-inspired systems. A very significant element of such systems is the nonlinear interaction between the involving fluids and structures. Although a common approach is to segregate the flow field information from the overall system and examine its dynamical properties, yet it would be very informative to incorporate the structural kinematics into these mathematical and computational frameworks to determine the levels of coupling between the fluid and structures. This approach would also enable us to segregate the contributions of fluid and structure towards the dynamics of their overall system. The only effort on this account found in literature is done by Goza and Colonius (2018) in which they considered a two-dimensional flow field around a flapping flag and analyzed its limit-cycle and chaotic dynamics.

In our current study, we present a computational framework to look for energetically strong modal decompositions for three-dimensional dynamical systems involving fluid-structure interactions. We employ the physiology of a Jack Fish the motion of which was recorded live by a high-speed photogrammetry system. First, utilizing proper orthogonal decomposition on the data set of its structural configurations enables us to propose a low-dimensional description of its complex flexible body kinematics. Next, we use proper orthogonal decomposition and dynamic mode decomposition approaches to investigate the fluid-structure interactive mechanics and explain the formations and productions of primary coherent fluid structures for two Reynolds numbers; 500 and 4000. We perform the modal analysis in a fluid-structure interaction framework for the flow only over the caudal fin because of the following two reasons, (1) the inclusion of a thick body structure in this computational framework would lead to spurious flow oscillations around and inside the body (Goza & Colonius, 2018; Menon & Mittal, 2020), and (2) the caudal fin is the primary thrust producing component for a Jack Fish as explained by Liu et al. (2017).

The manuscript is organized as follows. Section 2 explains our computational methodology to carry out numerical simulations using a sharp-interface immersed boundary method-based solver. It also provides details for our approach to conduct modal analyses of the kinematics of a Jack Fish and the fluid-structure interaction-based system composed of the membranous caudal fin and the vortical flow field around it. Next, we present our analyses and findings about the low-dimensional description of this highly nonlinear system using proper orthogonal decomposition and dynamic mode decomposition in section 3. Finally, we summarize and conclude our manuscript in section 4.

## 2. COMPUTATIONAL METHODOLOGY

In this section, we elucidate our computational methodology to handle the reconstruction of the physiological model of a Jack Fish and its kinematics. We also explain the numerical methodology based on a sharp interface immersed-boundary method to perform numerical simulations for flows over the caudal fin at Reynolds numbers 500 and 4000. Moreover, we summarize proper orthogonal decomposition and dynamic mode decomposition techniques and illustrate our strategy to set up snapshot data matrices for their further processing in order to extract the most dominant modal characteristics for the kinematics of a Jack Fish and the fluid-structure interaction based system of its caudal fin. Due to capturing the real fin motion and its incorporation in our computational solver, we argue that this whole system forms the basis of our claim about the interaction between the fluid flow and the structural oscillations in our present study. As explained in subsequent sections, the consistency in finding the Strouhal number as a constant for all the swimming speeds of the Jack Fish also supports this argument.

### I. Jack Fish Physiological Model and Kinematics

To reconstruct the geometry of a Jack Fish and its kinematics, we employ the data recorded and reported previously by (Liu, et al., 2017) to investigate the body-fin and fin-fin interaction during its steady swimming. Although the procedure to capture the fish motion and its physiology along with the statistical details have been covered in Ref. (Liu, et al., 2017), we present its salient points here as well for the sake of completeness. The current model is of Crevalle Jack (*Caranx hippos*) which is classified as a carangiform swimmer. Out of total 12 individuals of this class of fish with a mean total length $L = 0.338$ m and swimming at $1L/\text{sec}$ to $4L/\text{sec}$. It is important to highlight that their body kinematics did not change much with the increasing swimming speed. Currently used kinematic data was adopted from an individual fish with $L = 0.31$ m and swimming at $2\,L/\text{sec}$. The total height and width, normalized by $L$, of this fish are 0.286 and 0.144, respectively. The area, normalized by $L^2$, of the caudal fin is 0.023 and its normalized length is 0.244. The normalized height and length of the caudal fin are 0.315 and 0.244, respectively.

In this study, we consider the trunk and caudal fin only since these two components primarily contribute towards the kinematics and functionality of a fish. Its trunk is modeled as a solid body with a closed surface and the caudal fin is a membrane with zero thickness. Each surface is, then, represented by triangular mesh where the main body is composed of 11358 elements and 22712 nodes. The surface of the caudal fin has 1369 elements and 2560 nodes (see Figure 1c). The measured wavelength from the midline profiles is approximately $1.05L$ which is a characteristic of the carangiform swimming mode. The measured Strouhal number (St) for these recordings remain 0.30, where $\text{St} = f_E A/U$ with $f_E$ being the excitation/flapping frequency of the caudal fin, $A$ as the area of the caudal fin, and $U$ as the swimming speed.

## II. Numerical Solver

We perform three dimensional (3D) numerical simulations for flows over the oscillating caudal fin; a membranous structure, at Re = 500 and 4000. Following non-dimensional forms of the continuity and incompressible Navier-Stokes equations constitute the mathematical model for the fluid flow:

Continuity Equation:
$$\frac{\partial u_i}{\partial x_i} = 0$$

Navier-Stokes Equations:
$$\frac{\partial u_i}{\partial t} + \frac{\partial u_i u_j}{\partial x_j} = -\frac{\partial p}{\partial x_i} + \frac{1}{Re}\frac{\partial^2 u_i}{\partial x_j \partial x_j}$$

where the indices $\{i,j\} = \{1,2,3\}$, $x_i$ show Cartesian directions, the $u$ denote the Cartesian components of the fluid velocity, $p$ is the pressure, and Re represents the Reynolds number; defined as $Re = U_\infty L/\nu$. Here, $\nu$ indicates kinematic viscosity, $U_\infty$ stands for free-stream velocity, and $L$ is the entire body-length of the Jack Fish.

We solve the described governing model for fluid flow using a Cartesian grid-based sharp-interface immersed boundary method (Mittal, et al., 2008) where the spatial terms are discretized using a second-order central difference scheme and a fractional-step method is employed for time marching. This makes our solutions second-order accurate in both time and space. We utilize the Adams-Bashforth and implicit Crank-Nicolson schemes for the respective numerical approximation of convective and diffusive terms. The prescribed wavy kinematics is enforced as a boundary condition for the swimmers. We impose such conditions on immersed bodies through a ghost-cell procedure (Mittal, et al., 2008) that is suitable for both rigid and membranous body-structures. Further details of this solver and its employment to solve numerous bio-inspired fluid flow problems are available in Ref. (Liu, et al., 2017; Wang, et al., 2019; Han, et al., 2020; Wang, et al., 2020).

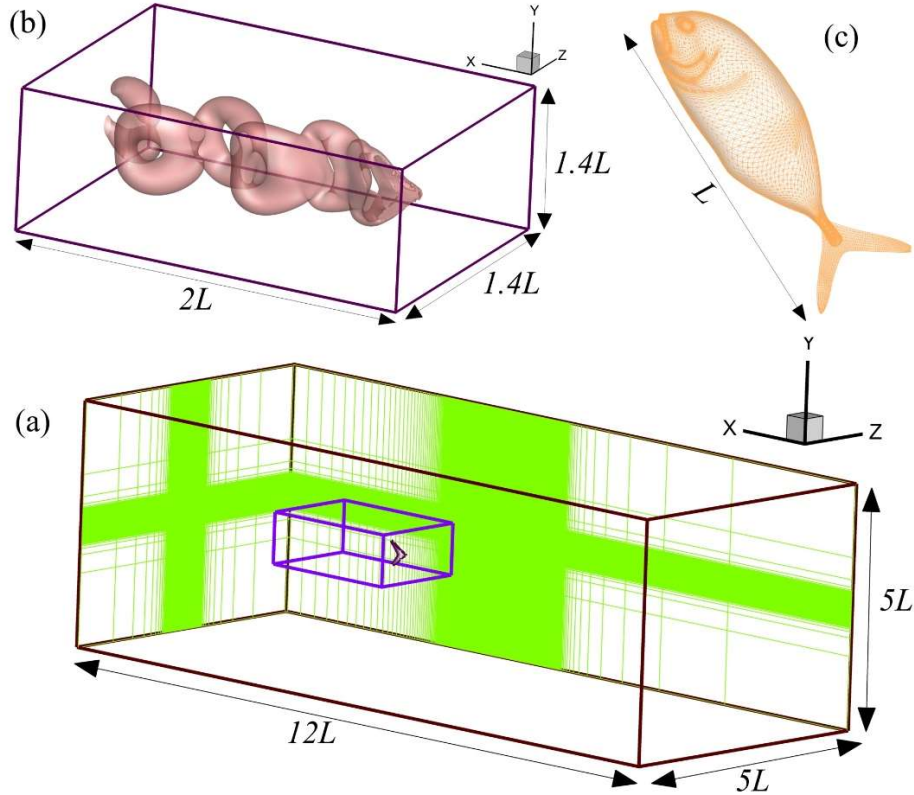

**Figure 1:** (a) Virtual tunnel for simulating flows over the caudal fin of a Jack Fish with its dimensions where the inner box shows the domain to extract data for modal decompositions, (b) a zoomed-in view of the inner domain covering the caudal fin and coherent flow structures in its wake, and (c) Jack fish body and caudal fin covered with a mesh to indicate the marker points to track their motion

Next, we employ Dirichlet boundary conditions for flow velocities at all the sides except the left one where Neuman conditions are used at the outflow boundary on the right side (see Figure 1a). The slices on the back and left boundaries show the regions with high mesh density in order to adequately resolve the flow features around the structure and its wake. The rectangular box in Figure 1b, encompassing the swimmer's body, shows the region of which we extract the data to perform our modal analysis. We use a mesh size $(N_x, N_y, N_z) = (385,129,161)$ for the complete flow domain, while the extracted domain for our further analysis has a mesh size $(n_x, n_y, n_z) = (234,123,153)$. It means that the total number of nodes in the entire flow domain and its extracted part are 7.99 million and 4.40 million, respectively.

### III. Proper Orthogonal Decomposition

Proper orthogonal decomposition technique provides us with a data analysis method focusing on extracting energy ranked modes to propose relevant mathematical models in order to describe the system dynamics with reduced dimensionality (Akhtar, et al., 2009). This strategy gives us optimal and orthonormal spatio-temporal modes of a dataset. POD modes and their associated useful information can be obtained by employing either eigenvalue decomposition of the covariance matrix of a dataset or by performing singular value decomposition (SVD) of the data matrix. In this data matrix, information about the states of a dynamical system is stored and arranged in particular patterns to further process it by utilizing these techniques.

In our present study, we perform the POD analysis through the SVD technique. The main focus here is to use modal decomposition methods for two purposes: (1) to extract significantly reduced-dimensional information for the complex wavy kinematics of a Jack Fish and (2) to propose a computational framework in order to perform modal analyses of fluid-structure interaction based systems and capture the most relevant information about both the structural motion and flow field. It is customary to exclude the time-averaged profiles of a dataset before applying POD. This practice makes it equivalent to principal component analysis (PCA) in the fields of imaging, video processing, and computer graphics (Perlibakas, 2004). To decompose the fish kinematics into its primary POD modes, we construct the following snapshot data matrix:

$$X = \begin{bmatrix} \xi_B^{t_1} & \xi_B^{t_2} & \cdots & \xi_B^{t_N} \\ \xi_{CF}^{t_1} & \xi_{CF}^{t_2} & \cdots & \xi_{CF}^{t_N} \\ \eta_B^{t_1} & \eta_B^{t_2} & \cdots & \eta_B^{t_N} \\ \eta_{CF}^{t_1} & \eta_{CF}^{t_2} & \cdots & \eta_{CF}^{t_N} \\ \zeta_B^{t_1} & \zeta_B^{t_2} & \cdots & \zeta_B^{t_N} \\ \zeta_{CF}^{t_1} & \zeta_{CF}^{t_2} & \cdots & \zeta_{CF}^{t_N} \end{bmatrix}_{(3N_B + 3N_{CF}) \times N_T}$$

where $\xi$, $\eta$, and $\zeta$ denote the displacements of each nodal point on the surface of the fish in $x$, $y$, and $z$ direction, respectively. The subscripts $B$ and $CF$ indicate the information belonging to the main body (trunk) and caudal fin, respectively. This snapshot matrix contains motion information of 48 time instants spanning one complete oscillation cycle of the Jack Fish. The SVD is formulated as:

$$X = U\Sigma V^T$$

where $U$ is a unitary matrix containing left eigenvectors of the snapshot data matrix $X$, $\Sigma$ is a diagonal matrix with positive numbered entries $\sigma_i$ termed as singular values and arranged in the descending order, i.e., $\sigma_1 \geq \sigma_2 \geq \sigma_3 \ldots \ldots \geq \sigma_N$, and $V$ is another unitary matrix. The eigenvalues ($\lambda$) can be computed by squaring $\sigma$ values. It is important to highlight that the columns of $U$ matrix give us the spatial distribution of POD modes, whereas $V$ contains the information about the temporal variations in these modes. Each column of the $V$ matrix provides us the temporal coefficients ($\alpha$) of the POD modes. To connect it with the traditional eigenvalue decomposition, $U$ and $V$ are the eigenvectors of covariance matrices $X^TX$ and $XX^T$, respectively. Conventionally, the sizes of $U$, $\Sigma$, and $V$ matrices are $(3N_B + 3N_{CF}) \times (3N_B + 3N_{CF})$, $(3N_B + 3N_{CF}) \times N_T$, and

$N_T \times N_T$, respectively, however, performing the "*economy*" SVD in MATLAB enables us to obtain and process $U$ and $\Sigma$ matrices with their respective sizes of $(3N_B + 3N_{CF}) \times N_T$ and $N_T \times N_T$ which reduces the computational burden to a large extent and prevents us from facing out-of-memory problems during numerical processing.

A reduced-order reconstruction for the system's kinematics or dynamics can be attained by using $U$, $\Sigma$, and $V$ matrices in the basic formulation of SVD. To obtain the temporal behavior (video) of a particular $i^{th}$ POD mode, we can make all the entries zero except its particular singular value $\sigma_i$ in the $\Sigma$ matrix. Thus, the mathematical form for this concept is;

$$X_i = U\Sigma_i V^T$$

In order to utilize this technique for a system based on fluid-structure interaction, we construct our snapshot matrix using the entries pertaining to the dynamical states of both the structure; caudal fin in this case, and fluid flow in the following form:

$$X = \begin{bmatrix} u^{t_1} & u^{t_2} & \dots & u^{t_N} \\ v^{t_1} & v^{t_2} & \dots & v^{t_N} \\ w^{t_1} & w^{t_2} & \dots & w^{t_N} \\ \xi_{CF}^{t_1} & \xi_{CF}^{t_2} & \dots & \xi_{CF}^{t_N} \\ \eta_{CF}^{t_1} & \eta_{CF}^{t_2} & \dots & \eta_{CF}^{t_N} \\ \zeta_{CF}^{t_1} & \zeta_{CF}^{t_2} & \dots & \zeta_{CF}^{t_N} \end{bmatrix}_{(3N_x+3N_y+3N_z+3N_{CF}) \times N_T}$$

Here, $u$, $v$, and $w$ are the Cartesian components of the fluid velocity, and each column represents a snapshot of the FSI system at one time instant.

## IV. Dynamic Mode Decomposition

Dynamic mode decomposition (DMD) provides a computational framework to extract a primary low-order description of a data set through its orthonormal modes in a temporal sense. The DMD modes are also approximations of the *Koopman Operator* which is a linear infinite dimensional operator representing a nonlinear dynamical system onto the *Hilbert Space* of the functions and states under consideration (Kutz, et al., 2016). It enables us to build a linear description of a complex dynamical system without losing its nonlinear characteristics. The sole idea is to construct a formulation of a dynamical system $x(t)$ such that $x(t_2) = Ax(t_1)$, $x(t_3) = Ax(t_2)$, and so on. In other words, we have $x(t_N) = A^{N-1} x(t_{N-1})$. This method computes DMD modes for the matrix $A$ by minimizing $\|x_k - Ax_{k-1}\|_2$, where the subscripts $k$ and $k-1$ are some time-instants.

For this purpose, we distribute the original snapshot data matrix $X$ into two submatrices $X_1$ and $X_2$, where $X_1 = [X^{t_1} \quad X^{t_2} \quad \dots \quad X^{t_{N-1}}]$ and $X_2 = [X^{t_2} \quad X^{t_3} \quad \dots \quad X^{t_N}]$. For further details on the algorithm, the readers are referred to references (Rowley, et al., 2009; Schmid, 2010; Schmid,

2011; Kutz, et al., 2016). In order to exploit the underlying nature of this data-driven technique, this algorithm indirectly solves for the DMD modes $\Phi$ of $A$ matrix. To reduce the computational cost incurred due to the large amount of data set, we, first, perform the POD and truncate the lowest energy-ranked modes to include the most relevant information in the DMD computations. The real and imaginary parts of the corresponding DMD eigenvalues; $\lambda_r$ and $\lambda_i$, respectively, denote the growth rate and frequency of DMD modes. Next, the associated angular frequencies having units rad/sec would be computed by $\omega = \log(\lambda)/\Delta t$, and its further division by $2\pi$ gives us the linear frequency in Hertz. Here, $\Delta t$ is the sampling time for attaining the snapshot data matrix. We obtain the approximate solution for the next time instants using the following form:

$$x(t) = \sum_{j=1}^{N} b_j \phi_j \exp(\omega_j t)$$

where $b_j$ is the initial amplitude, serving as the initial condition as well, of the $j^{th}$ mode.

## 3. RESULTS & DISCUSSION

Before discussing the modal analysis for the kinematics of a Jack Fish and the flow fields around its oscillating caudal fin at different Reynolds numbers, it is important to explain the temporal character and the frequency components of hydrodynamic forces on the caudal fin. For this purpose, we define the nondimensional hydrodynamic force components as $C = F/0.5\rho U^2 A_{CF}$, where $A_{CF}$ is the area of the caudal fin. Subscripts of $C$ represents the direction of each force component. Figure 2 presents temporal variations of the horizontal (drag/thrust) and lateral forces. We find that $C_X = C_D$ tend to change its instantaneous magnitude levels with a change in Re, and there exists a small change in its phase as well. Nonetheless, $C_Z$ shows almost similar patterns for both Reynolds numbers though higher Re causes a smaller increase in its positive and negative peak values. The spectral decomposition of horizontal, lateral, and sideways forces in Figure 3 reveals that the excitation frequency of the caudal fin $f_E$ is the most dominant frequency in the Fourier spectra of $C_Z$. However, both $C_D$ and $C_Y$ possess the frequency $2f_E$ as the strongest one with a smaller contribution from $f_E$ and its higher harmonics. These observations are consistent with those from flows over cylinders (Imtiaz & Akhtar, 2017) and flapping wings (Khalid, et al., 2015; Khalid, et al., 2018; Liang & Dong, 2015).

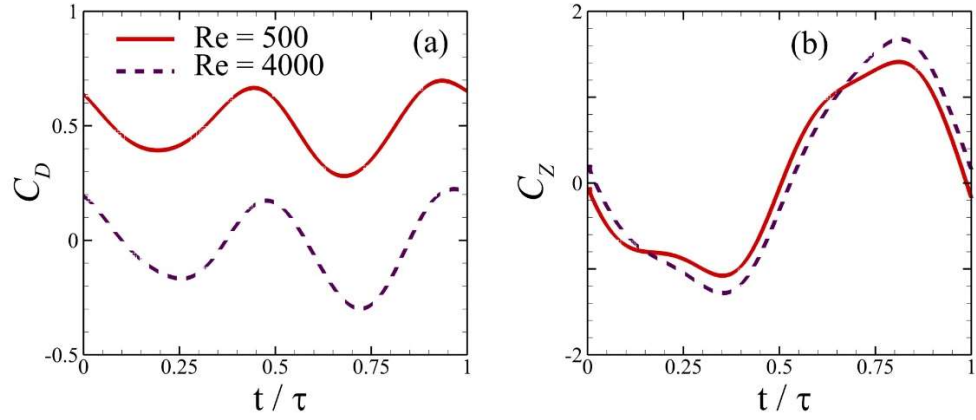

**Figure 2:** Time histories of hydrodynamic force coefficients in the horizontal and lateral directions

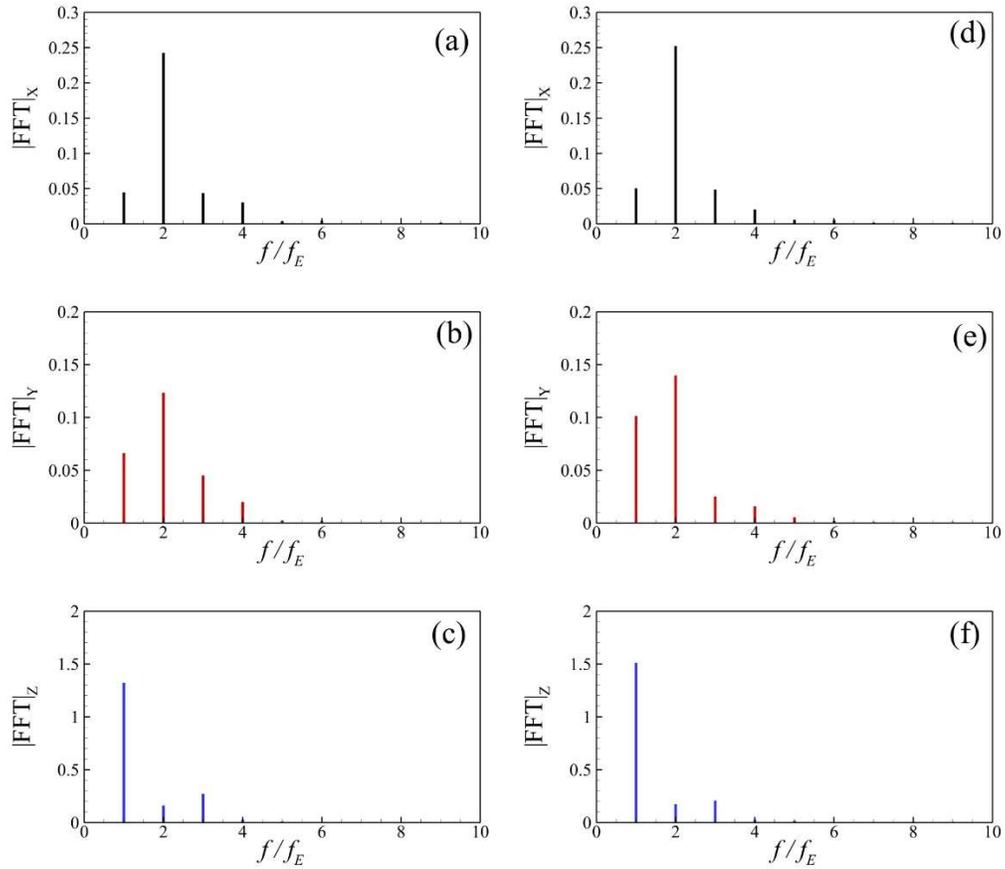

**Figure 3:** Fourier spectra of hydrodynamic force coefficients along the Cartesian axes, (a & d) drag/thrust force, (b & e) vertical force, and (c & f) lateral force, where the left and right columns show data for $Re = 500$ and 4000, respectively.

### a. Fish Kinematics

Due to the flexibility of their bodies, different species of fishes perform a complex wavy motion; usually known as undulation composed of travelling waves along their bodies. Using POD technique, we decompose its full-order kinematics into its spatially orthonormal modes. Figure 4 shows the POD eigenvalues normalized by their summation that represent the corresponding energy level of each mode. It is evident that the first POD mode constitutes approximately 79% of the energy, whereas the second POD mode has more than 20% energy. All the other modes carry almost zero energy levels. Thus, seemingly complex carangiform motion mainly comprises of only two primary modes.

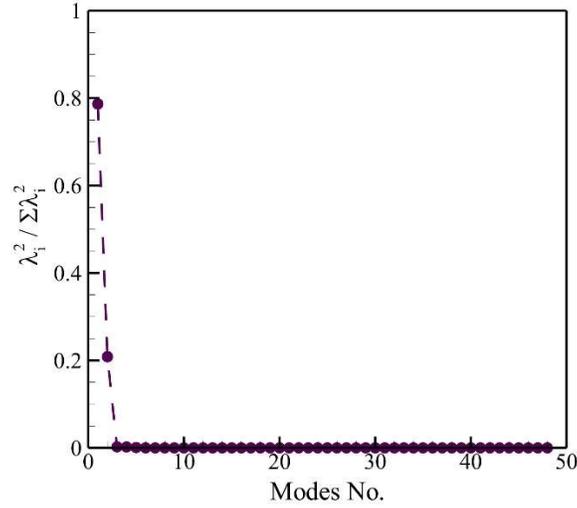

**Figure 4:** Energy levels of POD modes of a Jack Fish kinematics

To illustrate it further, we represent the instantaneous positioning of the fish trunk during its one oscillation cycle for the first two modes in Figure 5. In both the modal configurations, the posterior part of the body shows a greater amount of displacement from its equilibrium position that is a characteristic feature of carangiform and sub-carangiform swimming patterns. The first POD mode shows a standing wave like structure where the nodes and antinodes are evidently visible. The first node is located at 17% of the total body-length, whereas the second one is positioned at $0.59L$. The orientations of the trunk section of the second POD mode shows that the main body pitches about the $y$-axis passing through a point at $0.34L$. However, these configurations combined with those of the caudal fins give another standing wave along its length, where the first and second nodes are located at $\frac{x}{L} = 0.34L$ and $0.84L$. Moreover, looking at the caudal fin alone in its POD modes 2, we observe its pitching motion about the mid-points of its dorsal and ventral peripheries. Its formation in POD mode 1 exhibits a flapping motion; a combination of heaving and pitching. It is interesting to notice that both the POD modes demonstrate left-right asymmetry for the trunk section and the caudal fin. We observe prominent dorsal-ventral asymmetry for the caudal fin by comparing its orientations. It appears that the pitching angle of the ventral side of the caudal fin is lesser than that of its dorsal side. A careful look at the instantaneous configurations in Figure 5a and b reveal that there exists a phase angle of $\pi/2$ between the two POD modes. Hence, the entire undulatory kinematics of a Jack fish comes out to be the summation of its mean position and two standing waves moving with a phase of $\pi/2$.

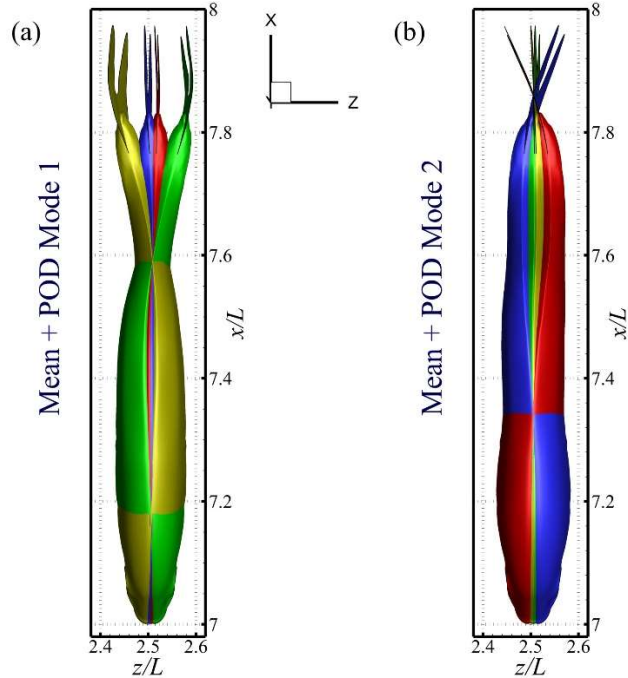

**Figure 5:** Modal configurations of the trunk and caudal fin of a Jack Fish where color coding is used to distinguish between the instantaneous positions: red ($\frac{t}{\tau} = 0.25$), green ($\frac{t}{\tau} = 0.50$), blue ($\frac{t}{\tau} = 0.75$), and yellow ($\frac{t}{\tau} = 1.00$)

### b. Flow Fields Analysis

Here, we perform numerical simulations for flows over the caudal fin of a Jack Fish at two Reynolds numbers; 500 and 4000. These two flow conditions are representatives of viscous (Re~$10^2$) and transition (Re~$10^3$) flow regimes.

Figure 6. As observed in the top-most row, there exist four distinct vortex tubes at Re = 500 and six tubal structures are present at Re = 4000. It seems that an increase in Re breaks the two vortex tubes on the dorsal side of the caudal fin, and the remaining two on the ventral side remain intact with a few signs of disruptions as they get developed in the downstream direction. Here, four tubes are elongated and the other newly developed two tubes formed due to the higher Re remain shorter. Because these structures traverse downstream at an inclination, they tend to diverge from each other. To elucidate the symmetry features of these flow fields, we plot contours of the Cartesian components of vorticity; $\omega_x$, $\omega_y$, and $\omega_z$, on surfaces normal to their corresponding axes. For Re = 500, the $x$-component of vorticity ($\omega_x$) show four distinct coherent structures reminiscent of the formation of four vortex tubes in the wake of the caudal fin. It is clear that $\omega_x$ demonstrates symmetry about the diagonal axis joining the two corners of the plane as drawn in Figure 6b.

In such a problem, there may exist four independent reflection symmetries, denoted as $S_X$, $S_Y$, $S_Z$ and $S_D$ with respect to the $x$, $y$, $z$-axes and a diagonal axis $y + z$. We use the following forms to mathematically define these symmetry characteristics.

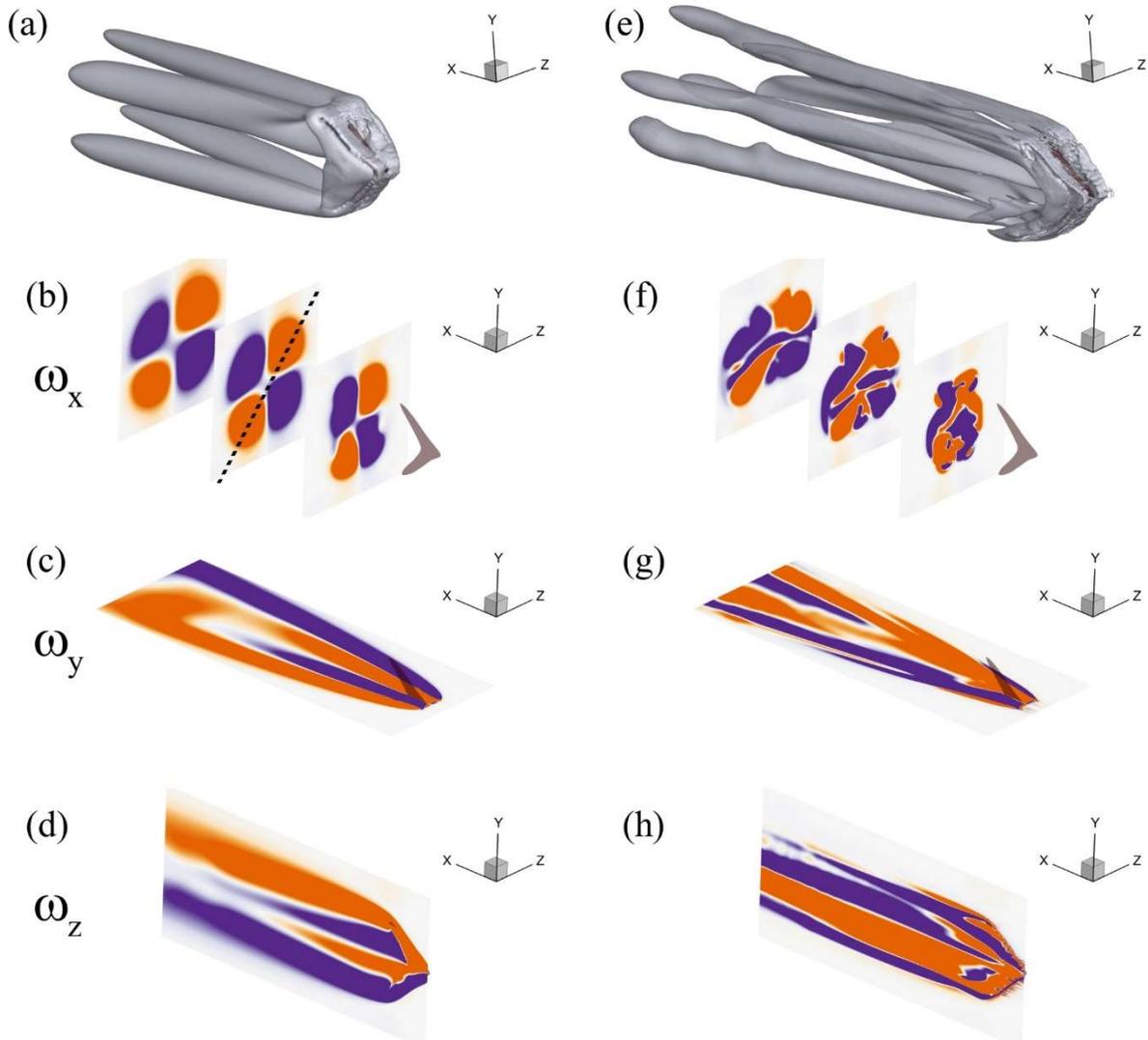

**Figure 6:** Mean flow fields and their symmetry properties using Cartesian components of vorticity where the left (a-d) and right (e-h) columns show data for $Re = 500$ and $4000$. The plots in the 1st row are drawn using Q-criterion. The 2nd, 3rd, and 4th rows represent data on the corresponding planes for $\omega_x$, $\omega_y$, and $\omega_z$, respectively.

$$S_X: \quad (x,y,z) \to (-x,y,z)$$
$$(u,v,w) \to (-u,v,w)$$
$$S_Y: \quad (x,y,z) \to (x,-y,z)$$
$$(u,v,w) \to (u,-v,w)$$
$$S_Z: \quad (x,y,z) \to (x,y,-z)$$
$$(u,v,w) \to (u,v,-w)$$
$$S_D: \quad (x,y,z) \to (x,z,y)$$
$$(u,v,w) \to (u,w,v)$$

Employing these terms, $\omega_x$ holds $S_D$ symmetry for the mean flow field at Re = 500, but we do not find any symmetry for $\omega_x$ at Re = 4000 with the existence of a few coherent structures here. Observing the contours of $\omega_y$ and $\omega_z$ on $xz$ and $xy$ planes, respectively, reveals the formation of four shear layers for the lower Re and six shear layers at the higher Re. Considering the three contour plots in Figure 6f, we come to know that the development of distinct coherent vortical structures should be carefully analyzed for complex flows as their orientations and characteristics may change as we move downstream. This may also result in the switching of symmetric and asymmetric flow features as would be explained later in the study.

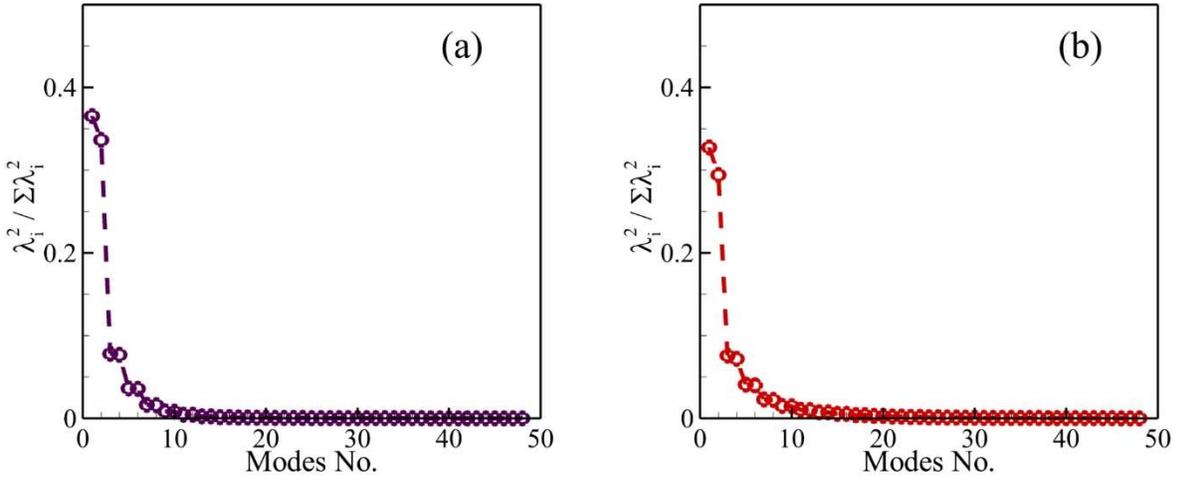

**Figure 7:** Squared eigenvalues normalized by their summation for (a) $Re = 500$ and (b) 4000

Performing POD for the snapshot data containing information for both the fluid and structural motion provides us with the knowledge about how much contribution each POD mode would have in this complex dynamical system. We present the energy levels of POD modes for our full FSI system through the squared and normalized eigenvalues in Figure 7 for Re = 500 and 4000. For the viscous flow regime, the first two modes contribute 36.5% and 33.6% energy to the overall system dynamics. The remaining modes exist in pairs due to the periodic oscillations of both the caudal fin and vortices in the wake. The same phenomenon was observed for flows over circular cylinders at very low Re previously (Taira, et al., 2020). Here, the first four modes would have more than 85% of the total energy of this dynamical system. Analyzing the data in Figure 7b for Re = 4000 also exhibits similar trends. Here, the POD modes 1 and 2, respectively, have 32.7% and 29.4% of the total energy. As expected, we need to include six POD modes to capture around 85% of the energy due to a smaller effect of viscosity under these conditions. Even though the viscous action is at a reduced level at the higher Re, we observe the making of pairs reflecting order to a large extent. Nevertheless, it is interesting to note that the structural elements in our data would only contribute to the first two POD modes for both the flow conditions because we do not see substantial oscillations of the caudal fin in the higher POD modes (see supplementary movies 5 and 6). It means that the caudal fin, as the oscillating structure, only contributes towards the development of the first two POD modes, and the oscillatory patterns in the higher POD modes find their origin in the fluid dynamics only.

Now, we present the POD modes of our FSI system at Re = 500 and 4000 in Figure 8 (a-c) and when we see the higher POD modes here, although the formation and presence of coherent flow structures are evident there as well.

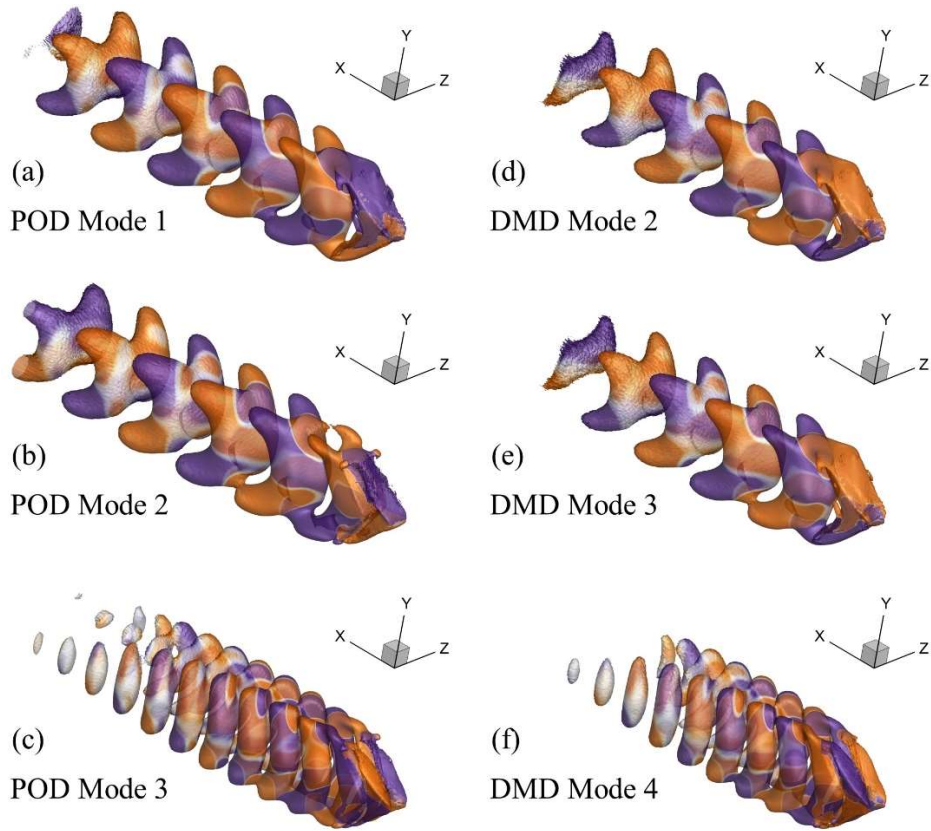

**Figure 8:** Most dominant oscillatory POD and DMD modes for our FSI system working at $Re = 500$

Figure *9* (a-c), respectively. For the lower Reynolds number conditions, the flow structures tend to form a shape of a "headless panda" (quadfurcated shape) with four small arms extended in the downstream direction. Nevertheless, these arms like structures vanish in the higher POD modes and we see only planar structures aligned closely with each other. For Re = 4000, these features adopt hairpin-like shapes as presented in Figure 9, but, these flow features lose their distinct shape

In

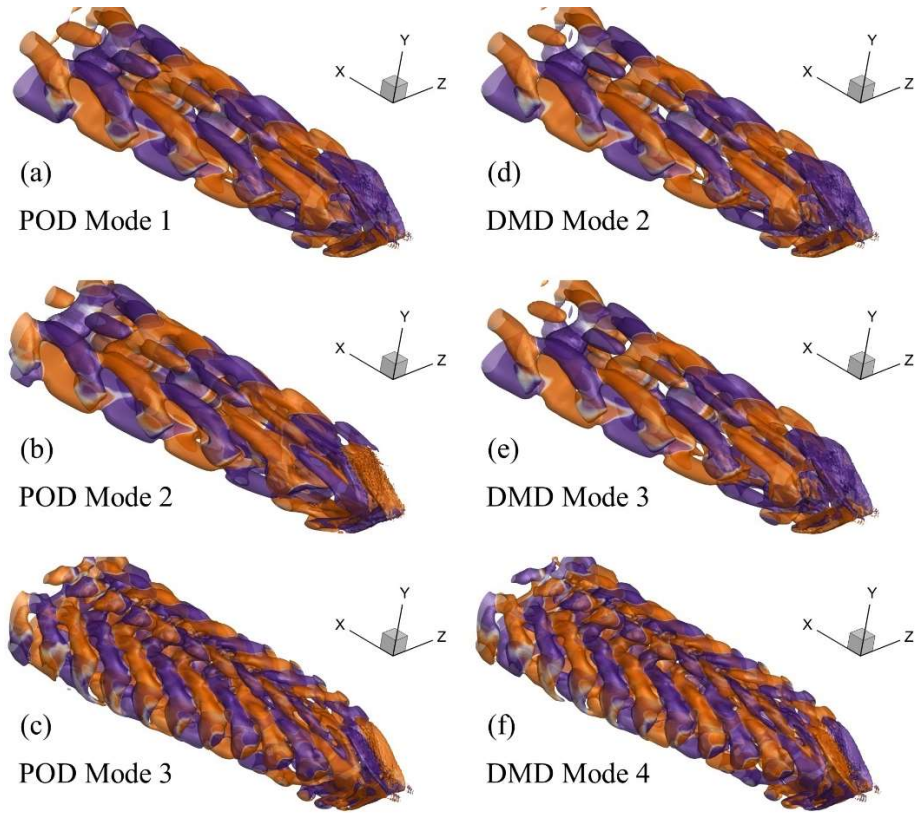

**Figure 9:** Most dominant oscillatory POD and DMD modes for our FSI system working at $Re = 4000$

, we provide the symmetry characteristics of the first 8 POD modes for each flow condition. We discover that lateral component of vorticity ($\omega_z$) always show asymmetry, whereas the other two components; $\omega_x$ and $\omega_y$, would show variations in their properties. It is also important to note that each pair of POD modes would possess similar characteristics, despite the complex motion of the caudal fin and the vortex patterns in its wake. While moving the $xy$, $yz$, and $xz$ planes along their corresponding normal axes, we find that the symmetry properties of these dynamical systems do not remain consistent throughout the wake, rather, they may momentarily switch their states with some asymmetric pattern to regain symmetry afterwards. We highlight those conditions in
. An important feature of our analysis is that, for a few significant POD modes at Re = 4000, $\omega_y$ would exhibit symmetric coherent patterns about the axis parallel to the $xz$-plane and cutting it into two halves only when the plane lies in the middle of the domain. Thus, it is of utmost importance that extreme care should be taken while performing such analyses using experimental techniques where middle planes are usually selected to fin the traits of coherent structures. Another salient observation is the increasing number of asymmetric flow patterns of the POD modes at the higher Re.

**Table 1:** Symmetry properties of Mean and POD Modes of our FSI systems

| Reynolds No. | POD Mode Number | $\phi_{\omega_x}$ | $\phi_{\omega_y}$ | $\phi_{\omega_z}$ |
|---|---|---|---|---|
| 500 | Mean | $S_d$ | Asymmetry (shear layer, no coherent structures) | Asymmetry (shear layer, no coherent structures) |
| | 1 | $S_y$ | $S_x$ | Asymmetric |
| | 2 | $S_y$ | $S_x$ | Asymmetric |
| | 3 | $S_d$ – Asymmetry Switching | Asymmetric | Asymmetric |
| | 4 | $S_d$ – Asymmetry Switching | Asymmetric | Asymmetric |
| | 5 | $S_y$ | $S_x$ | Asymmetric |
| | 6 | $S_y$ | $S_x$ | Asymmetric |
| | 7 | Asymmetric | Asymmetric | Asymmetric |
| | 8 | Asymmetric | Asymmetric | Asymmetric |
| 4000 | Mean | Asymmetric | Asymmetry (shear layer, no coherent structures) | Asymmetry (shear layer, no coherent structures) |
| | 1 | $S_y$ – Asymmetry Switching | $S_x$ (on the mid-plane only) – Asymmetry Switching | Asymmetric |
| | 2 | $S_y$ – Asymmetry Switching | $S_x$ (on the mid-plane only) – Asymmetry Switching | Asymmetric |
| | 3 | Asymmetric | Asymmetric | Asymmetric |
| | 4 | Asymmetric | Asymmetric | Asymmetric |
| | 5 | $S_y$ – Asymmetry Switching | $S_x$ (on the mid-plane only) – Asymmetry Switching | Asymmetric |
| | 6 | $S_y$ – Asymmetry Switching | $S_x$ (on the mid-plane only) – Asymmetry Switching | Asymmetric |
| | 7 | Asymmetry | Asymmetry | Asymmetry |
| | 8 | Asymmetry | Asymmetry | Asymmetry |

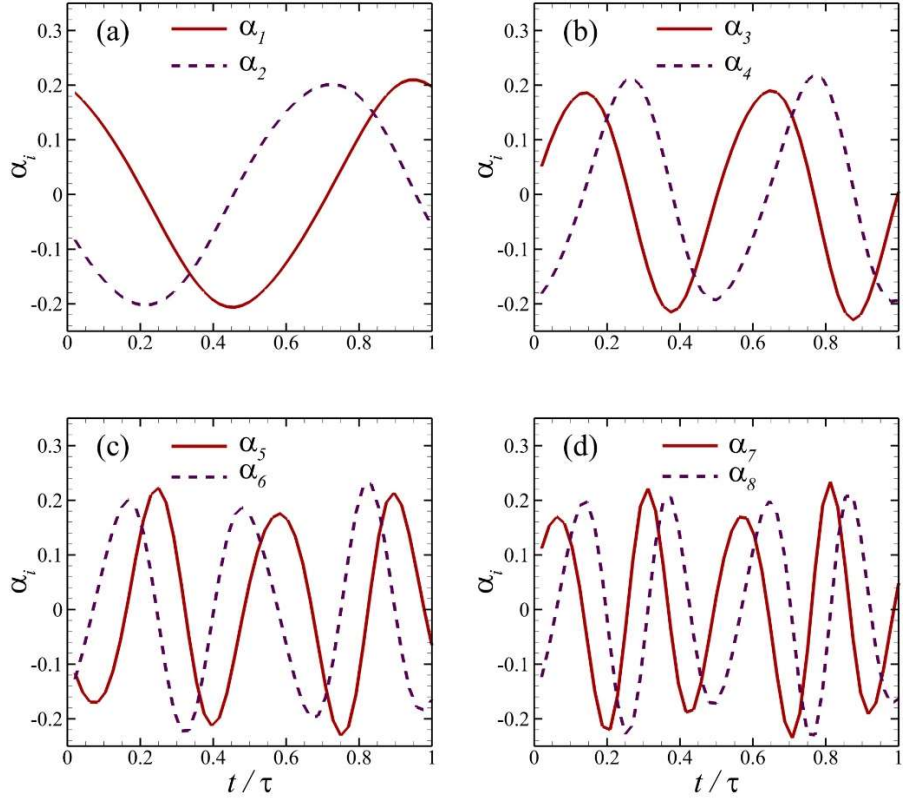

**Figure 10:** Temporal coefficients for the POD modes of our FSI system at $Re = 500$

Next, we plot the temporal coefficients of the first eight POD modes for our FSI system at Re = 500 in Figure 10. These coefficients would remain the same in their trends and magnitude levels at Re = 4000, not shown here for the sake of brevity. Which modes contribute to the production of respective hydrodynamic forces can be understood by comparing the dominant frequency contents of each temporal coefficient with those of hydrodynamic forces on the caudal fin presented in Figure 3. It is evident that our POD modes 1 and 2 would have more contribution towards the production of lateral hydrodynamic force, $F_z$, whereas the thrust production is more associated with POD modes 3 and 4. We argue that the temporal coefficients for the POD modes 1 and 2 undergo one oscillation cycle in one time-period that shows their most dominant frequency to be equal to the excitation frequency. We observe the same pattern for the oscillations of the lateral force in Figure 2b. Comparing the time-histories of the POD modes 3 and 4 with those of the drag/thrust force in Figure 2a, it is clear that these parameters have $2f_E$ as the most dominant frequency. Here, the higher modes carry components due to a combination of the fundamental frequency with its second harmonic. It is important to reiterate here that an increase in Re does not change the temporal features of the POD modes and the associated variations are only reflected in their spatial characteristics.

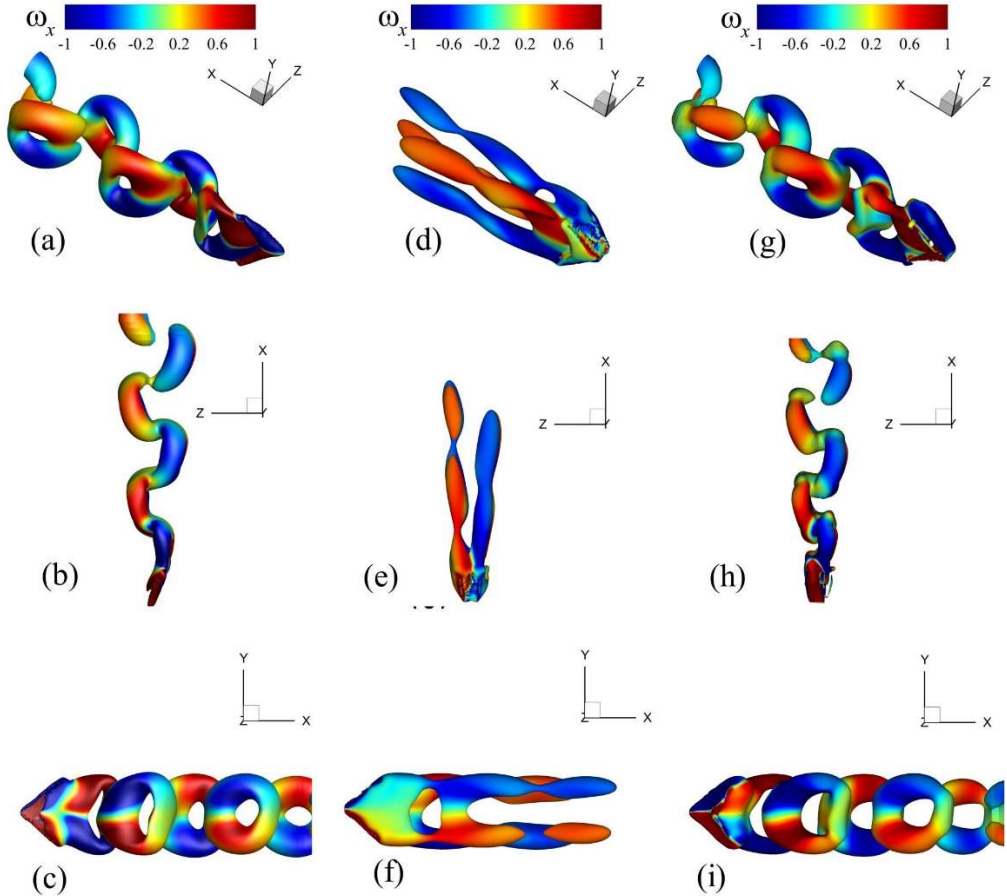

**Figure 11:** Reconstruction of FSI dynamics for a flow over the caudal fin of a Jack Fish at $Re = 500$ using mean features and the POD modes, where the left, middle, and right columns show the full-order fluid-structure dynamics, the addition of the mean states with the POD mode 1, and the addition of the mean states with the POD modes 1 and 2, respectively. Each row shows the perspective, top, and side view of the whole flow and structural domains.

Now, we reconstruct the flow and structural dynamics using the time-averaged structural state and flow field and stepwise additions of POD modes to illustrate the contributions of each POD mode towards the production of oscillations in the wake. Using critical values through the Q-criterion here, we provide vortex visualizations where the addition of POD mode 1 with the mean flow and structural fields explains the role of this first oscillatory mode to cause the breaking of the vortex tubes (see Figure 6a) in the primary direction of the flow. It is evident that the POD mode 1 would play the key role in determining the wavelength of the coherent structures in the wake. Moreover, bringing the POD mode 2 into this system would cause the emergence of connecting legs of the vortices to produce coherent flow patterns. For Re = 500, only the mean fields added with the first two POD modes would be sufficient to reconstruct the intricate details of the FSI dynamics although the contribution of these two modes is limited to 70% of the total energy of this system.

Furthermore, we witness that the POD mode 1 computed for Re = 4000 plays the same role in breaking the vortex tubes (not shown here) in the mean flow field (see Figure 6e). Nevertheless, the addition of another mode (POD mode 2) in this case would not reconstruct the major features of the vortex dynamics as it does for our viscous flow conditions. To develop meaningful connections between the broken parts of the vortex tubes, we need to add at least the first four POD modes into the time-averaged field. The POD modes 2, 3, and 4 play their parts for the development and growth of fluidic connections in the lateral and sideways directions to produce significant coherent structures in the wake, as shown by the perspective, top, and side views of the flow domain in Figure 12.

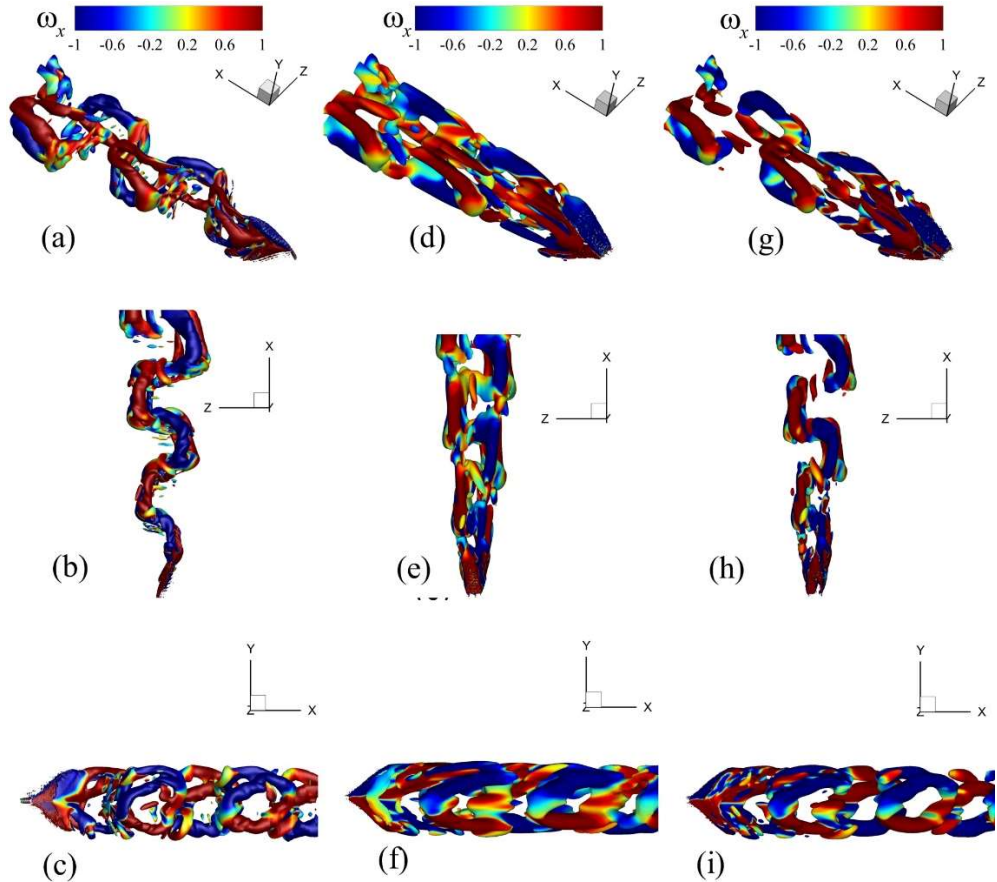

**Figure 12:** Reconstruction of FSI dynamics for a flow over the caudal fin of a Jack Fish at Re = 4000 using mean features and the POD modes, where the left, middle, and right columns show the full-order fluid-structure dynamics, the addition of the mean states with the first three POD modes, and the addition of the mean states with the first four POD modes, respectively. Each row shows the perspective, top, and side view of the whole flow and structural domains.

Next, we perform dynamical mode decomposition for the full FSI systems constructed from the structural motion of the caudal fin and the flow field around it. An important objective of this analysis is to segregate the frequency components from each other in this bio-inspired dynamical system where the modes are temporally orthonormal, whereas the temporal coefficient of each POD mode contains more than one frequency. In Figure 13, we plot the angular frequencies (rad/sec) of the DMD modes that are computed by taking the logarithm of Ritz values (DMD eigenvalues) and dividing it by the sampling time interval. Unlike POD, these eigenvalues are not arranged in the descending order, and one needs to carefully determine the significant DMD modes and their associated parameters. A parameter to rank these modes is the amplitude of the DMD modes (Kutz, et al., 2016). These DMD eigenvalues come up with their complex conjugates because we process the real-valued data here. The Ritz values existing on a unit circle indicate neutrally stable modes, whereas those inside the circle and outside its periphery show decaying and unstable DMD modes, respectively. Here, the real value of an angular frequency ($\omega_r$) on the left side of the vertical axis in Figure 13 reflect the stability of that particular DMD mode. All the DMD modes with their $\omega_r = 0$ are neutrally stable. We neither find any DMD mode with $\omega_r > 0$ for Re = 500 nor 4000. The most dominant mode, in both the cases, indicated by red circles in Figure 13 are the mean DMD modes, also referred to as DMD mode 1. These modes have $\omega_i = 0$ that shows their non-oscillatory character. For Re = 500, the first three strongest oscillatory DMD modes have $\omega_r = 0$ which means that they do not decay with time. All the other modes with angular frequencies have negative $\omega_r$ and would decay as we progress in time. This parabolic arrangement of modal frequencies has also been observed previously by Schmid et al. (Schmid, 2010; Schmid, 2011). In the case of Re = 4000, only DMD mode 2 has $\omega_r = 0$. All the remaining modes show the decaying character. Under both the flow conditions, the DMD mode 2 has the excitation frequency of the caudal fin, whereas DMD modes 3 and 4 have frequencies equal to $2f_E$ and $3f_E$, respectively.

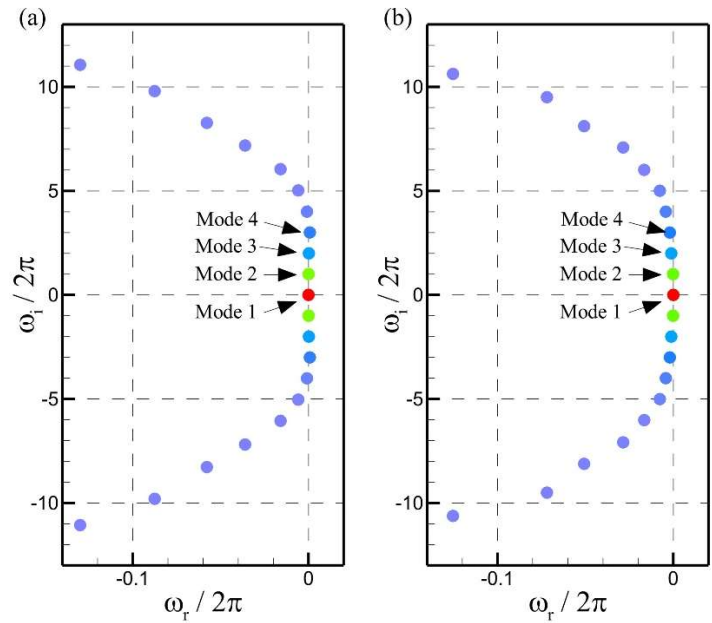

**Figure 13:** Real and imaginary components of the angular frequencies computed from the Ritz values for (a) $Re = 500$ and (b) 4000

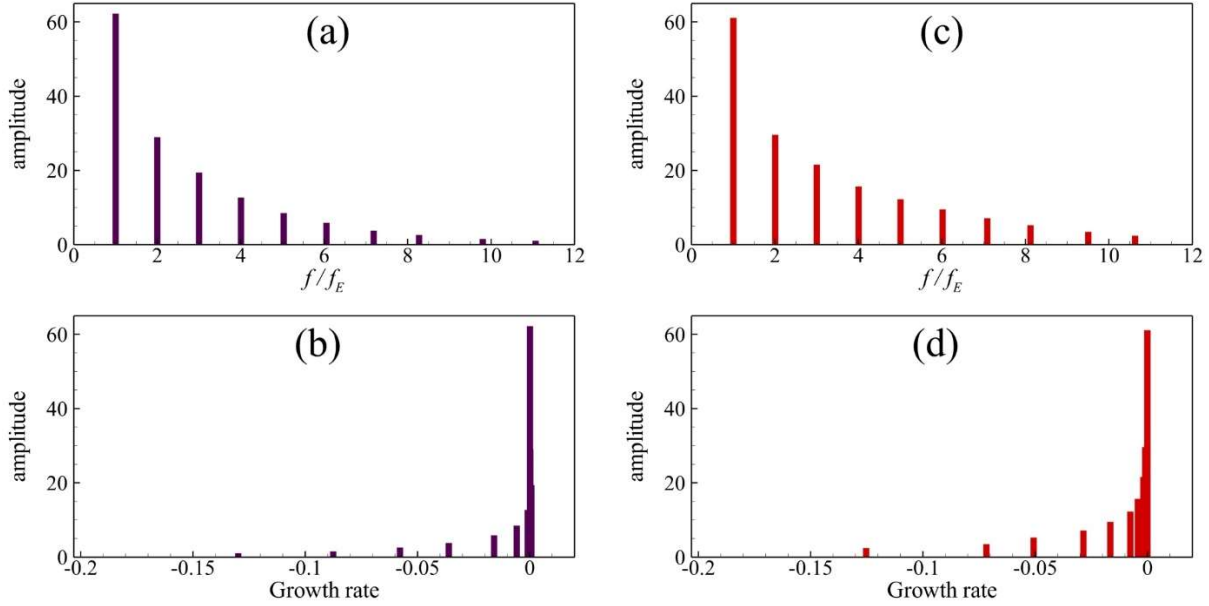

**Figure 14:** DMD modal frequencies and growth rates vs modal amplitudes for $Re = 500$ (a and b) and $Re = 4000$ (c and d)

To illustrate more on the strength of these DMD modes, we plot frequencies, normalized by the excitation frequency of the caudal fin, and growth rates for each modal component versus modal amplitude in Figure 14 for both Reynolds numbers. Here, we only show the information for oscillatory components. We determine that the amplitudes of these frequencies would decay asymptotically (see Figure 14a and c), and no DMD mode has a growth rate greater than zero. This later feature exhibits the stable character of the dynamics of this fluid-structure interaction-based system. It is important to mention that the strength of higher DMD modes is more in the case of Re = 4000 as compared to those at Re = 500. Modal distributions in Figure 14d shows the cluster of the strongest DMD modes near the zero value on the horizontal axis. Our analysis establishes that the DMD mode 3 would contribute more towards the production of thrust force on the caudal fin, whereas the second DMD mode would have a greater effect on the lateral force.

As also mention by Taira et al. (2020), POD and DMD modes are usually similar for periodic flows. In our present study, we observe this phenomenon not only for highly viscous conditions but also at a relatively higher Re~$10^3$. We illustrate these observations by comparing the POD and DMD modes presented in Figure 8 and Figure *9* for Re = 500 and 4000, respectively. For the lower Re, the DMD flow structures also appear as planar elements with the four extended arms in the downstream direction, whereas, their shapes turn out to be hairpin-like for the higher Re.

## 4. SUMMARY AND CONCLUSIONS

In this work, we perform modal decomposition of the kinematics of a Jack Fish that is a carangiform swimmer. We find that its complex undulatory motion is mainly composed of two dominant modes which represent standing waves with different locations of nodes and antinodes along the fish's body and its caudal fin. These two modes are sufficient to present the wavy kinematics of this natural aquatic swimmer and this information can be used to build a low-order model for further studies. Then, we perform numerical simulations for flows over the membranous caudal fin of the Jack Fish using our immersed boundary method-based computational solver. We employ a large amount of data from this complex FSI system to extract dominant modes using proper orthogonal and dynamic mode decomposition techniques. Proper orthogonal decomposition modes provide us with spatially orthonormal structures, whereas the other technique decomposes the entire information about the structural and flow fields into orthonormal frequency components. Each POD mode carries more than one frequency, but each DMD mode has only one frequency. We find that only two modes are sufficient to reconstruct the structural and flow dynamics at the lower Reynolds numbers. However, we need to bring in a greater number of modes to capture essential dynamical features of the flow field at $Re = 4000$. It means that high modal oscillations occur only due to the fluid side, and not the structural motion. We also illustrate the symmetry properties using vorticity components plotted on their respective normal planes in the wakes and reveal that there exists diagonal symmetry for certain POD modes. We also emphasize that these symmetry properties may be switched to asymmetric patterns when their corresponding planes are moved along their normal axes. We find similarities in respective POD and DMD modes for both flow conditions. The coherent structures take exhibited quadfurcated shapes with four extended arms in the downstream directions in both POD and DMD modes, but we see hairpin-like structures for the flow at the greater Reynolds number. Even these systems are representatives of intense fluid-structure interactions, yet we reveal the formation of stable and neutrally stable DMD modes here.